# Causal Inference in Disease Spread across a Heterogeneous Social System

Minkyoung Kim[1], Dean Paini[2], and Raja Jurdak[1]

[1]Data61, Commonwealth Scientific and Industrial Research Organisation (CSIRO), Queensland, Australia
[2]Health & Biosecurity, CSIRO, Canberra, Australia

***Background.*** Diffusion processes are governed by external triggers and internal dynamics in complex systems. Timely and cost-effective control of infectious disease spread critically relies on uncovering the underlying diffusion mechanisms, which is challenging due to invisible causality between events and their time-evolving intensity.

***Methods.*** We infer causal relationships between infections and quantify the reflexivity of a meta-population, the level of feedback on event occurrences by its internal dynamics (likelihood of a regional outbreak triggered by previous cases). These are enabled by our new proposed model, the Latent Influence Point Process (LIPP) which models disease spread by incorporating macro-level internal dynamics of meta-populations based on human mobility.

***Results.*** We analyse 15-year dengue cases in Queensland, Australia. From our causal inference, outbreaks are more likely driven by statewide global diffusion over time, leading to complex behavior of disease spread. In terms of reflexivity, precursory growth and symmetric decline in populous regions is attributed to slow but persistent feedback on preceding outbreaks via inter-group dynamics, while abrupt growth but sharp decline in peripheral areas is led by rapid but inconstant feedback via intra-group dynamics.

***Conclusions.*** Our proposed model reveals probabilistic causal relationships between discrete events based on intra- and inter-group dynamics and also covers direct and indirect diffusion processes (contact-based and vector-borne disease transmissions).

Diffusion processes in the real world often produce non-Poisson distributed event sequences, where inter-event times are highly clustered in the short term but separated by long term inactivity [3, 35]. Examples are observed in both human and natural activities such as resharing microblogs in online social media [6, 17, 39], citing scholarly publications [15, 18, 30, 28], a high incidence of crime along hotspots [21, 29], and aftershock sequences near the seismic center [32]. These all imply that an event occurrence is likely triggered by preceding events in cascades of different scales, and the timing of discontinuous events conveys information of underlying diffusion mechanisms.

Uncovering such feedback mechanisms between preceding and triggered events has drawn significant attention from a wide range of scientific communities [2, 9, 14], since it helps predict diffusion trends and establish cost effective strategies for the promotion or restriction of the diffusion process. When it comes to epidemics, an accurate understanding of underlying dynamics is crucial for the timely control of infectious disease spread. However, uncovering disease dynamics is very challenging due to often unknown causal relationships between events and limited information of private social networks, unlike explicit citation relationships in online social media or in academic publications. Moreover, large international and domestic travel volumes have increased the uncertainty of infection pathways. Thus, the quantification of exogenous and endogenous effects is essential to overcome the challenges and understand emergent bursts of infections, which has been however largely neglected in epidemic studies [8, 5, 25].

In this study, we propose the Latent Influence Point Process model (LIPP) which infers causality between events and quantifies the reflexivity of a meta-population, *i.e.*, the level of feedback on event occurrences by its internal dynamics across a heterogeneous social system. The proposed model incorporates three major counter-balancing factors: (1) exogenous influence covering environmental heterogeneity of target regions, (2) endogenous influence by incorporating human mobility across multiple regions, and (3) time decay effect on disease spread.

We first simulate our model so that we can generate synthetic data based on diverse parameter settings that mimic real data. We then evaluate the model performance of parameter recovery in order to examine how well our model parameters are recovered from data, which is important to validate our interpretations of diffusion dynamics with real data. As a case study, we investigate dengue spread in Queensland, Australia during a period of 15 years (2002–2016), provided by Queensland Health. From our causal inference, we find that dengue outbreaks become more globally interconnected across multiple regions over time, leading to more complex behavior of disease spread. In terms of reflexivity, precursory growth and symmetric decline of outbreaks in metropolitan or populated regions is attributed to slow but persistent feedback on preceding outbreaks via inter-group dynamics, while abrupt growth but sharp decline in remote or peripheral regions is driven by rapid but inconstant feedback (frequent outbreaks during an intensive period) via intra-group dynamics. Similar diffusion trends between regions reflect synchronous feedback mechanisms of regional social systems. That is, the timing and size of event bursts in each region are comparable, which is likely driven by human mobility.

To the best of our knowledge, this study is the first to infer infection routes of a vector-borne disease, by modeling internal dynamics of meta-populations driven by human mobility as multivariate stochastic processes. In this way, our proposed model uncovers self and mutual excitation nature of disease spread across a heterogeneous social system with rich context. We expect that our proposed model can be readily applied to a wide range of intra- and inter-group diffusion processes in social and natural systems at a macro level and can infer probabilistic causality between discrete events.

## METHODS

In this section, we explain our data collection and propose a new diffusion framework which quantifies exogenous and endogenous dynamics in disease spread by incorporating external heterogeneity and internal dynamics of meta-populations at a macro level.

## Data Collection

**Dengue outbreak data.** We investigate dengue outbreaks in Queensland, Australia from 2002 to 2016, provided by Queensland Health. The dengue virus is a mosquito-borne viral disease transmitted among humans by mosquito vectors, whose outbreak risk is rapidly increasing worldwide [37]. The data contains records of anonymised infected individuals such as onset dates, residence postcodes, and acquisition places if available. For understanding cross-regional infections at a macro level, we categorized residence postcodes into 15 regions which correspond to the statistical areal level 4 (SA4) defined by the Australian Statistical Geography Standard (ASGS). Based on selected target regions, we create an event sequence as a tuple consisting of occurrence time and region identity.

**Table 1** Notation.

| Symbol | Description |
|---|---|
| $R$ | a set of regions |
| $r_n$ | region where $n$-th infection has occurred, $r_n \in R$ |
| $t_n$ | time of $n$-th infection, $t_n \in [0, T]$ |
| $N$ | total number of infections |
| $D$ | entire history of infections, $D = \{(t_n, r_n)\}_{n=1}^{N}$ |
| $T$ | an observation period |
| $\eta_r$ | intrinsic environmental risk of region $r$, $\eta_r > 0$ |
| $\xi_r$ | endogenous influence of region $r$, $\xi_r > 0$ |
| $\rho_r^k$ | directional human mobility from region $k$ to $r$ such that $\sum_{k \in R} \rho_r^k = 1$ |
| $\varphi_r$ | time decay exponent of region $r$, $\varphi_r > 0$ |
| $\lambda_r(t)$ | infection intensity of region $r$ at time $t$ |
| $\lambda_r^k(t)$ | infection intensity of region $r$ influenced by region $k$ at time $t$ |

**Human mobility.** We incorporate human mobility as topological heterogeneity across multiple regions, which reflects macro-level internal dynamics in a social system. In order to obtain structural connectivity between regions, we employ three different types of travel data such as International Visitor Survey [33], National Visitor Survey [34], and Twitter ( [41] section S1).

## Proposed Model

**Background.** We consider a Hawkes process [11] as our fundamental diffusion framework, since it is a non-Markovian extension of the Poisson process and thus realizes the clustering of events in the real world. The general univariate Hawkes process is defined with an intensity function,

$$\lambda(t) = \mu + \sum_{t_i < t} g(t - t_i) , \quad (1)$$

where the first term $\mu$ represents a background intensity by external influence. The second term characterizes endogenous feedback by weighting the influences of past events on future events. That is, the intensity of event occurrences is dependent on the history of preceding events.

Figure 1 shows the embodiment of a Hawkes process to a disease-spread scenario. As shown in Figure 1(a), disease infections are represented as a single arrival process, which disregards infections by self and mutual excitations (intra- and inter-region disease transmission) in Figure 1(b). As discussed earlier, such cross-regional outbreaks are accelerated by human mobility. In this context, the objective of our framework is to model bursty behavior (clustered in time and space) of disease outbreaks across meta-populations by incorporating human mobility as topological pathways in a heterogeneous social system.

**Model.** We now propose the Latent Influence Point Process model (LIPP) which incorporates the exogeneity and endogeneity of a social system as major components for realizing bursty diffusion processes in the real world. Based on inputs of a spatial and temporal event sequence and cross-regional human mobility, our model aims to quantify the reflexivity of a social system using inferred model parameters.

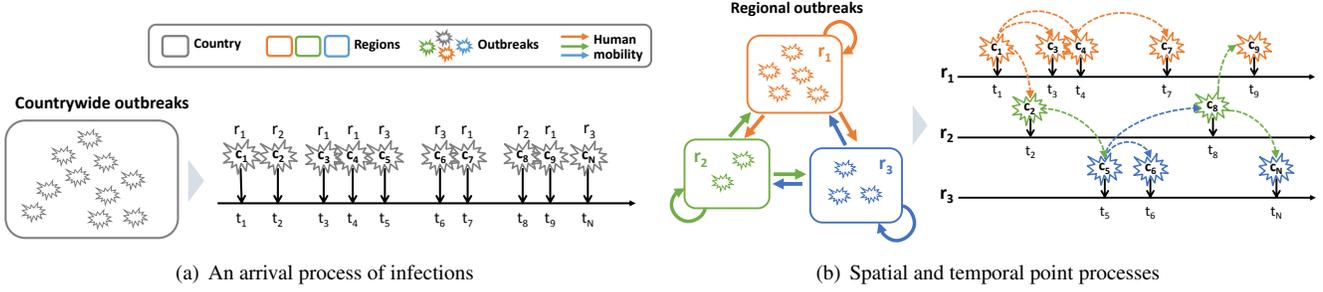

(a) An arrival process of infections

(b) Spatial and temporal point processes

**Figure 1.** Disease spread as point processes. (a) nationwide outbreaks of an infectious disease over time ($c_i$: $i$-th contagion at time $t_i$ in region $r_j$, and $N$: total outbreaks during an observation period). (b) nationwide outbreaks in (a) can be decomposed into timelines of each region, which can be represented as spatial and temporal point processes. Regions are color-coded, thick colored arrows represent human mobility between regions, and dashed arrows indicate hidden infection trajectories via social interactions.

Suppose that we observe an event sequence $\boldsymbol{D}$ consisting of $N$ spatio-temporal events in a set of regions $\boldsymbol{R}$ during an observation time period $[0, T]$. Here, each event is represented by a pair of its occurrence time $0 < t_n < T$ and region $r_n \in \boldsymbol{R}$ as a tuple such that $\boldsymbol{D} = \{(t_n, r_n)\}_{n=1}^N$, and the events are sorted by their time moments. As shown in Figure 1(b), we consider multiple timelines separated by event occurrence regions. For each region $r$, the history of events consists of two different types of event sequences, $\boldsymbol{D}_r^0$ and $\boldsymbol{D}_r^k$, influenced by external sources $0 \notin \boldsymbol{R}$ and triggered by preceding events at neighboring regions $k \in \boldsymbol{R}$, respectively. Given the whole event sequence $\boldsymbol{D}_r = \cup_{k \in \boldsymbol{R}_+} \boldsymbol{D}_r^k, \boldsymbol{R}_+ \equiv \boldsymbol{R} \cup \{0\}$, we assume that each event sequence is generated by a Poisson process. Thus, region $r$'s intensity function of time $\lambda_r(t)$ is defined based on the superposition property of Poisson processes [7] as

$$\lambda_r(t) = \lambda_r^0 + \sum_{k \in \boldsymbol{R}} \lambda_r^k(t) \ . \quad (2)$$

That is, we consider doubly stochastic point processes defined by $\lambda_r^k$ as our diffusion framework for the realization of intra- and inter-region disease transmissions, which corresponds to multi-dimensional Hawkes processes.

We incorporate three major counter-balancing components into our framework: (1) exogenous influence covering environmental heterogeneity of target regions, (2) endogenous influence attributed to macro-level interactions between meta-populations, and (3) time decay effect with an exponential memory kernel. Details are discussed as below.

**Exogenous intensity.** In region $r$, events can occur independently of a previous event history due to external influence. This is modeled with a Poisson process with a background intensity,

$$\lambda_r^0 = \eta_r \rho_r^0 \ , \quad (3)$$

Where $\eta_r > 0$ denotes disease-specific environmental heterogeneity of region $r$ (environmental infectiousness of a target disease). That is, region $r$ has an intrinsic environmental risk that does not change much over time, such as average temperature and humidity, annual precipitation, and distributions of disease vectors (*e.g.*, mosquito vectors in dengue), and thus some areas are more likely to cause disease outbreaks than others. The second term, $\rho_r^0 > 0$ represents the probability that an infection occurs in region $r$ by external exposures such that $\sum_{r \in \boldsymbol{R}} \rho_r^0 = 1$. For instance, international visitors from virus endemic regions outside $\boldsymbol{R}$ (*i.e.*, region 0) trigger local outbreaks in region $r$.

**Endogenous intensity.** Contrary to exogenous infections, internal dynamics in a social system drive bursts of events going through interactions between individuals over social networks, so it is also called internal influence. Our model incorporates cross-regional human mobility as macro-level endogenous effects on diffusion, and the intensity brought by mutual excitations across multiple regions is defined as

$$\lambda_r^k(t) = \sum_{t_i < t} \zeta(k) \rho_r^k \phi(t - t_i) \ , \quad (4)$$

where $\zeta(k) = 1 + \delta(k) \cdot \xi_k$, and $\delta(k)$ represents the indicator function of mosquito-vector presence at region $k \in \boldsymbol{R}$. Here, $\xi_k > 0$ represents the latent influence (infectiousness) of region $k$ on other regions due in part to population density and social interactivity in that region. The second term $\rho_r^k > 0$ represents the strength of directed connectivity from region $k$ to $r$ based on human mobility patterns such that $\sum_{r \in \boldsymbol{R}} \rho_r^k = 1$. For instance, the center of a city has a larger floating population than neighboring suburbs (connected by commute or travel routes), and thus it more likely triggers further infections in its neighboring regions. That is, $\zeta(k)$ weights human mobility $\rho_r^k$ from region $k$ to $r$, based on $k$'s latent influence $\xi_k$. Finally, the third term in Eq. (4) captures the time relaxation function for reflecting the effect of time decay on the likelihood of spread. Here, we consider an exponential memory kernel such that $\phi(t - t_i) = \exp(-\varphi_k (t - t_i))$ where $\varphi_k > 0$ indicates the time decay parameter for region $k$, *i.e.*, the level of infectiousness decay in region $k$.

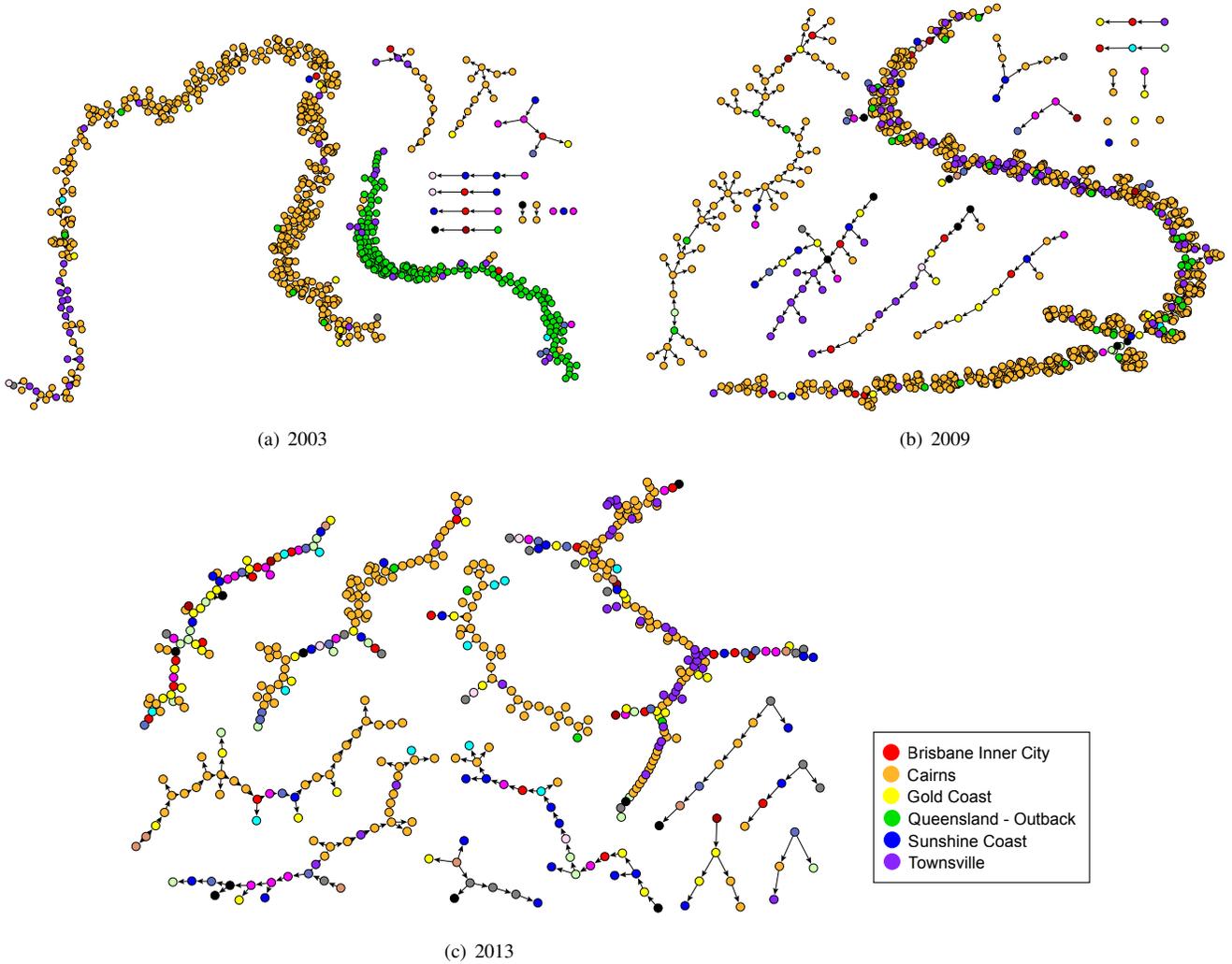

**Figure 2.** Event networks based on inferred causal relationships between events (dengue cases) for the selected years, (a) 2003, (b) 2009, and (c) 2013, when the largest dengue outbreaks had occurred. Each node and link indicate a dengue case and causality between preceding and triggered events with the highest probability, respectively. Nodes are color-coded by different regions, and six region names with the largest outbreaks are only presented for brevity. See Figure S2 for the comparison of event networks colored by provided outbreak IDs.

## Experimental Design

For causal inference, we introduce the latent index variables $Z = \{z_i\}_{i=1}^{N}$ consisting of event indicators each of which has triggered the $i$-th event, since causal relationships between events are unknown as discussed in Figure 1(b) (dashed lines). By using the Expectation-Maximization (EM) algorithm [40], we infer our model parameters, $\eta_r$, $\xi_r$, and $\varphi_r$ for each region $r \in \mathbf{R}$, representing environmental heterogeneity, latent influence, and time decay exponent, respectively ( [41] section S2). We simulate our proposed model so that we can generate synthetic data as the ground truth and evaluate the model performance. As evaluation metrics, parameter recovery errors are examined with respect to the mean absolute percentage error (MAPE). We also evaluate relative strengths between estimated parameters, which is important to validate our subsequent interpretations of underlying diffusion processes with real data ( [41] section S3). After verifying the proposed model performance with synthetic data, we conduct experiments on real data from dengue outbreaks and interpret underlying dynamics in the next section.

## RESULTS

In this section, we conduct experiments on real data and interpret estimated model parameters based on the verification of parameter recovery with synthetic data ( [41] section S3). For this experimentation, we set the length of observation time window as one year in order to examine time-evolving internal dynamics with a fine-grained time resolution.

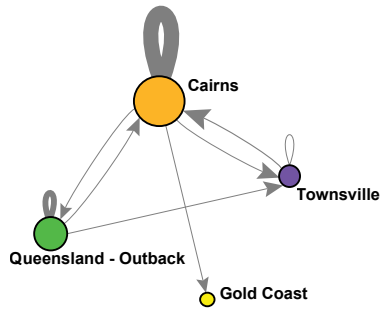
(a) 2003

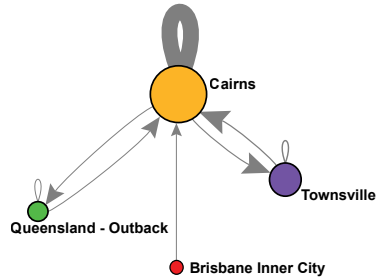
(b) 2009

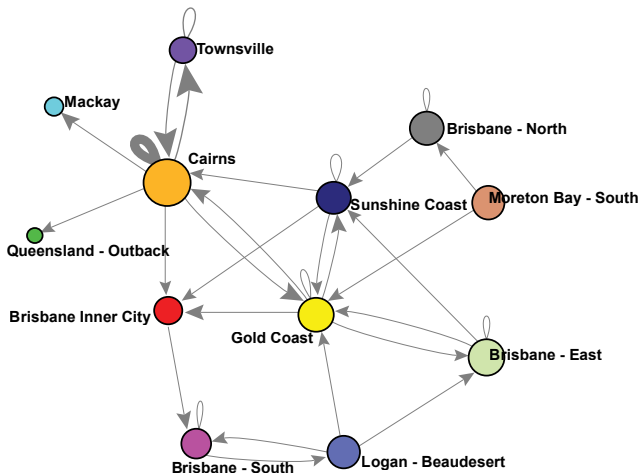
(c) 2013

**Figure 3.** Macro-level cross-regional causal relationships based on event networks in Figure 2. Node are color-coded as Figure 2, labels denote region names in Queensland, and node sizes are proportional to the number of dengue cases occurred. Arrow heads face influenced regions, and the width of a causal link indicates the strength of influence. Only links triggering more than 1% of dengue cases are presented for brevity. Self-loops and links represent self and mutual excitation for each.

## Causal Inference

As discussed earlier and shown in Figure 1(b) and Figure 4, causal relationships between events are unknown, so we infer the probability that each preceding event has triggered a current event by using EM algorithm ( [41] section S2).

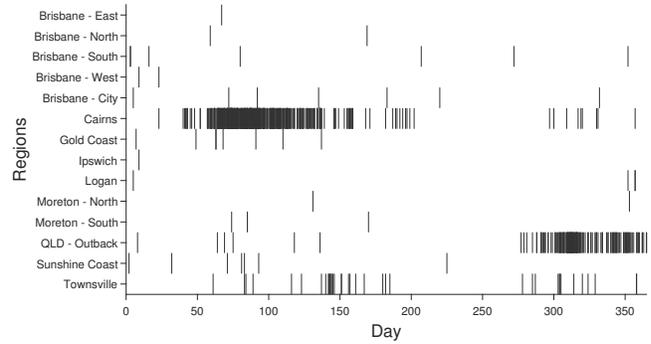
(a) 2003

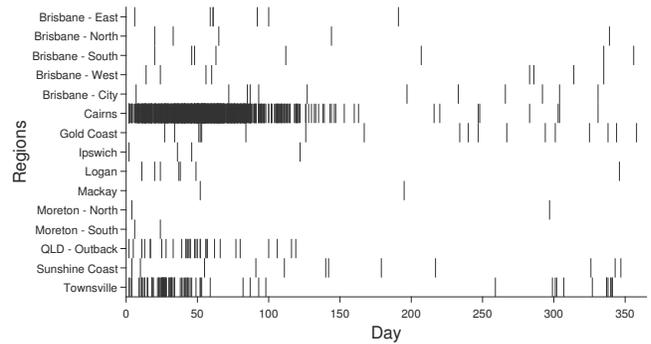
(b) 2009

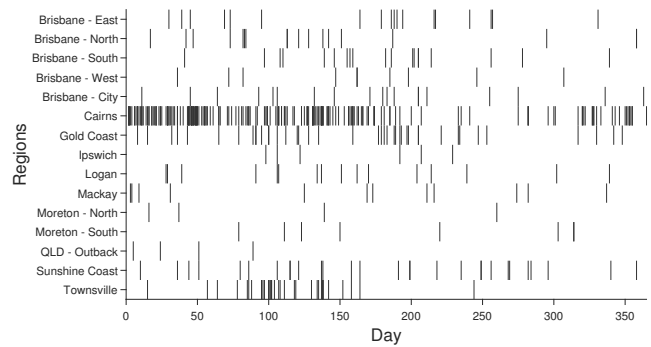
(c) 2013

**Figure 4.** Raster plots of event occurrences (dengue cases) during a year ($x$-axis) in each region ($y$-axis) for the selected years as in Figure 2 and 3.

**Time-evolving causality.** Figure 2 shows the results from our causal inference between dengue cases for the 3 years with the largest outbreaks during the 15-year period (see Figure S3 for the distribution of dengue cases over years), and nodes are color-coded by different regions. As the figure shows, earlier dengue outbreaks tend to be more locally clustered, but over the years they become globally interconnected across regions, leading to more complex behavior of infectious disease spread. Additionally, event networks are also colored-coded by provided out-break IDs in Figure S2, where a single outbreak ID is considered to be a collection of cascading (or ongoing) local transmission initiated by the same index case. As this figure shows, dengue cases are well clustered according to the outbreak IDs, reflecting the effectiveness of our causal inference.

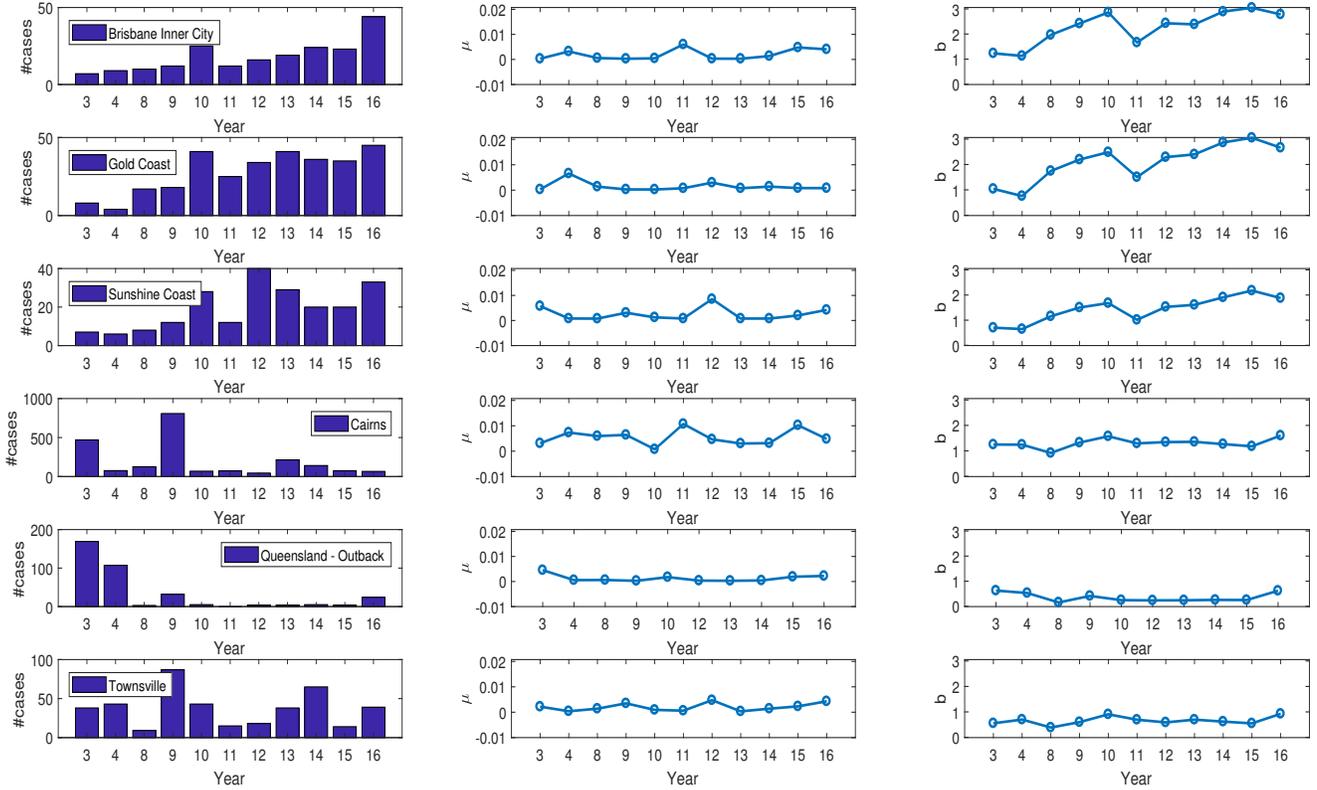

**Figure 5.** Reflexivity of regional social systems. The plots in the first column show the distributions of dengue outbreaks in each region during our observation period, while the second and third columns illustrate the level of exogeneity $\mu$ and endogeneity $b$ in dengue spread over time. Six regions are selectively chosen with the largest number of cases among 15 regions, and only years are considered with more than 150 dengue cases. For brevity, years are presented with the last two digits of the 2000s.

Figure 3 illustrates the summary of causal relationships between regions in Figure 2 at a macro level. As shown in the figure, spread of dengue becomes more far-reaching across Queensland over time. This is consistent with event raster plots in Figure 4, exhibiting increasing dengue outbreaks all over the regions throughout the year in 2013, compared with local outbreaks during an intensive period in 2003 and 2009. Such spatial expansion of infectious diseases can be attributed to the increase of human mobility and travel volumes [5, 36].

## Reflexivity of a Regional Social System in Disease Spread

The Hawkes process generalizes the nonhomogeneous Poisson process by allowing the self-exciting nature via preceding events, as discussed in Eq. (1). The linearity of the conditional intensity $\lambda(t)$ helps quantify the level of exogeneity and enogeneity in diffusion processes and align with a branching process consisting of triggers and descendants [10]. The branching ratio $b$ represents the average number of triggered events per initiating event and is defined as

$$b = \int_0^\infty g(t)dt \ . \quad (5)$$

**Reflexivity of Regional Social Systems.** Our proposed model extends multi-dimensional Hawkes process. From each regional intensity $\lambda_r(t)$ in Eq. (2), the background intensity and branching ratio correspond to $\mu = \eta_r \rho_r^0$ and $b = \sum_{k \in \mathbf{R}} \frac{1}{\varphi_k} \zeta(k) \rho_r^k$. Accordingly, we quantify the level of exogeneity $\mu$ and endogeneity $b$ for all target regions based on the estimated parameter values with our real data. Figure 5 summarizes these quantifications for regions with the largest dengue cases during the 15-year period and for years with more than 150 cases.

**Precursory vs. Abrupt Growth.** As Figure 5 shows, in general the background intensity $\mu$ hardly changes, while the branching ratio $b$ increases over time in metropolitan or populated ares such as Brisbane Inner City (BIC), Gold Coast (GC), and Sunshine Coast (SC) relatively to remote or peripheral areas such as Cairns, Outback, and Townsville. As shown in the first column of the figure, these two groups of regions also exhibit different patterns in the growth of dengue cases: precursory and gradual growth in populous regions (*i.e.*, BIC, GC, and SC) versus abrupt rise in peripheral regions (*i.e.*, Cairns, Outback, and Townsville) in 2003 and 2009, showing a split in behavior. Additionally, mosquito vectors are presented in the three peripheral regions, while absent in the other populous areas.

**Intra- and Inter-group Dynamics.** Precursory growth in the major population centers is likely due to high reachability from statewide regions, *i.e.*, a high probability of importation of infected individuals. However, the absence of mosquito vectors in these regions allow no more excitations by previous outbreaks, leading to symmetric decline. In other words, dengue outbreaks in these populous regions are driven by strong but unsustainable inter-group dynamics. On the other hand, abrupt growth but sharp decline in the peripheral regions is attributed to rapid but inconstant excitations via mosquito vector transmissions. This strong but unstable intra-group dynamics is possibly affected by time-varying vector density and visitor volumes. That is, non-uniformly distributed mosquito vectors statewide, unbalanced human mobility between regions, and time-varying visitor volumes compositely lead to such behavioral split.

**Synchronous Feedback Mechanisms.** Interestingly, BIC and GC exhibit the similar growth patterns and reflexivity of a regional social system, which implies that endogenous feedback mechanisms are synchronous. This is likely attributed to human mobility patterns, as shown in Figure S1. In other words, the large volumes of external visitors to these regions as well as large reciprocal fluxes between them more likely drive self and mutual excitations. Our proposed model can capture such emergent phenomena by incorporating endogenous effects of cross-regional human mobility on diffusion processes.

## DISCUSSION

The spread of infectious diseases leads to form event clusters in both space and time. Such spatio-temporal events are well realized by a point process due to its flexible consideration of lasting impact of bursty behaviors rather than a current snapshot [15], and thus it is widely used as a mathematical tool in diverse research areas [10,31]. In this context, we proposed a new model, LIPP which generalizes multi-dimensional Hawkes processes by incorporating macro-level internal dynamics of meta-populations, driven by human mobility.

**General extension.** By introducing latent indicator variables for triggering events, our proposed model can infer a causal link between preceding and triggered events with the highest probability and also quantify the level of exogeneity and endogeneity in disease spread without the need of prior knowledge on social network structures. By considering the presence of virus vectors in a target region ($\zeta(k)$ in Eq. (4)), LIPP extends processes of contact-based transmissions into vector-borne virus infections. These aspects increase the applicability of our proposed model to a wide range of intra- and inter-group diffusion processes in social and natural systems at a macro level.

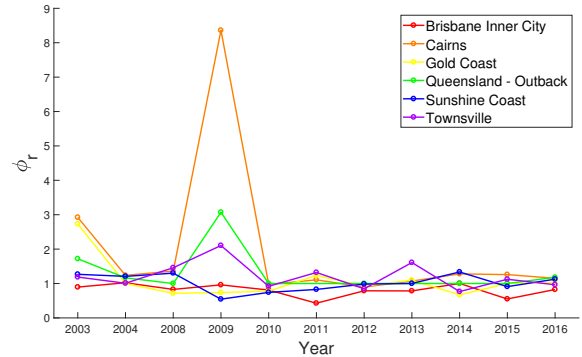

**Figure 6.** Estimated time decay effect ($\varphi_r$) of the proposed model for the six regions in Figure 5.

**Causal inference.** In terms of causal relationships between events, tracking infection routes often depends on time-consuming site investigations or a survey of infected patients' travel routes. Based on such efforts and expert knowledge, related outbreaks are manually linked as shown in Figure S2, but a considerable proportion of cases are left unknown or possibly misidentified. Here, our causal inference can provide investigators or experts with initial reference of probabilistic transmission paths for their efficient tracking and timely control of infectious diseases, reducing response time and cost under resource constraints.

**Interplay between growth rate and time decay.** For understanding the origin of a burst, the interplay between external shock and internal dynamics in complex systems has also been of great interest across disciplines [14,23,26]. Regarding the interplays between the growth rate of dengue cases and reflexivity of a meta-population in each region, Figure 6 shows estimated time decay effect ($\varphi_r$ in Eq. (4)) of our proposed model for the regions and years as in Figure 5. As shown in these figures, fast feedback (rapid time decay) in Cairns, Outback and Townsville for the years of 2003 and 2009 comes up with the sharp increase of outbreaks triggered by preceding events. On the other hand, populous regions such as BIC, GC, and SC show slow but persistent feedback, leading to gradual increase of dengue cases.

**Human mobility and complex behavior.** As discussed in Figure 3 and 4, dengue outbreaks in Queensland become more statewide more recently, leading to more complex causal relationships between events. This study considers human mobility as the vital medium of disease transmission at a macro level, leading to such spatial expansion of outbreaks. However, micro-level investigations, such as targeting sub-regions and analyzing socio-economic factors, also need to be considered for a holistic understanding of underlying diffusion mechanisms, which is an interesting direction for future work.

## Notes

**Author contributions.** M. K. proposed a model, preprocessed data, performed experimentations, and interpreted results, and wrote the manuscript. D. P. jointly conceived the study, guided the analysis, and edited the manuscript. R. J. jointly conceived and designed the study, guided the analysis, interpreted the results, and edited the manuscript. All authors read and approved the final manuscript.

**Acknowledgments.** We would like to express our gratitude to Cassie Jansen and Fiona May at Queensland Health for providing dengue outbreak data and for valuable discussions and to Tourism Research Australia for providing visitor survey data and helpful information.

# Supplementary Material: Causal Inference in Disease Spread across a Heterogeneous Social System

# Contents

# S1 Data Collection for Human Mobility

As discussed in our proposed model, we incorporate human mobility as topological heterogeneity across multiple regions, which reflects macro-level internal dynamics in a heterogeneous social system. In order to obtain structural connectivity between regions, we employ three different types of travel datasets such as International Visitor Survey (IVS), National Visitor Survey (NVS), and Twitter, as described in Table S1. IVS [1] and NVS [2] are conducted by Computer Assisted Personal Interviewing (CAPI) in the departure lounges of the international airports in Australia. These datasets consist of traveler information (*e.g.*, home country or residence, gender, age group), travel time periods, trajectories of visiting locations in Australia. Finally, geo-tagged Twitter data provides daily movement patterns in contrast to irregular movements of tourists. Thus, combining these different types of human mobility datasets enables us to construct more exhaustive and complementary population fluxes across regions.

We construct a network whose nodes and links correspond to regions in a SA4 level and normalized fluxes between regions, respectively. The constructed network is accordingly shown in Figure S1, where the size of nodes represents external fluxes ($\rho_r^0$) from outside of Queensland to each region and the width of links is proportional to internal fluxes ($\rho_r^k$) between regions.

# S2 Inference

We infer our model parameters by using the Expectation-Maximization (EM) algorithm. As discussed in our model formulation, our model parameters are $\eta_r$, $\xi_r$, and $\varphi_r$ for each region $r \in \boldsymbol{R}$, representing environmental heterogeneity, latent influence, and time decay exponent, respectively.

Table S1: Description of International Visitor Survey, National Visitor Survey, and geo-tagged Twitter data.

| Data | Collection Period | #Visits | #Persons |
|---|---|---|---|
| IVS | 2005 - 2015 | 1,274,903 | 442,445 |
| NVS | 1998 - 2015 | 751,184 | 588,323 |
| Twitter | 2015 | 925,945 | 79,271 |

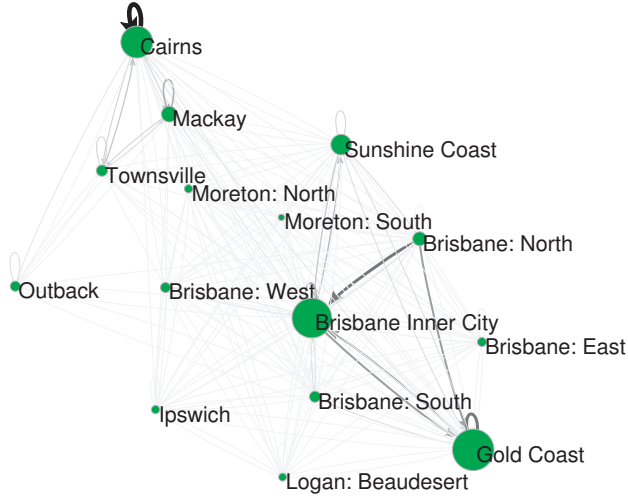

Figure S1: Estimated human mobility by using travel data in Table S1. The node size and link width are proportional to external fluxes from outside of Queensland and internal fluxes between regions, respectively. Self-loops and straight links represent intra- and inter-region human mobility.

**Likelihood.** According to the property that inter-event times generated by a Poisson process follow the exponential distributions, we first calculate the likelihood of observing an event sequence $\boldsymbol{D} = \{\boldsymbol{D}_r\}_{r \in \boldsymbol{R}} = \{(t_n, r_n)\}_{n=1}^N$ during an observation period $[0, T]$ as

$$p(\boldsymbol{D} \mid \boldsymbol{\eta}, \boldsymbol{\xi}, \varphi) = \exp\left(-\int_0^T \sum_{r \in R} \lambda_r(t) dt\right) \prod_{i=1}^N \lambda_{r_i}(t_i) \ , \tag{1}$$

where $\boldsymbol{\eta} = \{\eta_r\}_{r \in \boldsymbol{R}}$ and $\boldsymbol{\xi} = \{\xi_r\}_{r \in \boldsymbol{R}}$ denote all target regions' environmental heterogeneity and latent influence respectively.

We introduce the latent index variables $\boldsymbol{Z} = \{\boldsymbol{z}_i\}_{i=1}^N$ consisting of event indicators each of which has triggered the $i$-th event, since causal relationships between events are unknown as discussed in Figure 1(b) (dashed lines) in Major Article. Each latent variable $\boldsymbol{z}_i = \{z_{ij}\}_{j=0}^{i-1}$ is represented as an $i$-dimensional binary vector (e.g., $\boldsymbol{z}_i = [0, 1, 0, ..., 0]^\top \in \mathbb{R}^i$ when $i$-th event is triggered by the second event). We impose a Bernoulli distribution prior as

$$p(\boldsymbol{z}_i) = \prod_{j=0}^{i-1} (\pi_{ij})^{z_{ij}} \ , \ \boldsymbol{\pi}_i = \{\pi_{ij}\}_{j=0}^{i-1} \ , \tag{2}$$

where $\pi_{ij} > 0, \sum_{j=0}^{i-1} \pi_{ij} = 1$, and $\boldsymbol{\pi}_i$ is a set of probabilities of selecting preceding event $j \in \{0, ..., i-1\}$ which has triggered event $i$. Note that $j$ starts from 0 representing external influence so that a prior covers an initial event.

The likelihood in Eq. (1) with the latent variables is

$$p(\boldsymbol{D}, \boldsymbol{Z} \mid \boldsymbol{\Theta}, \boldsymbol{\Pi}) = p(\boldsymbol{D} \mid \boldsymbol{Z}, \boldsymbol{\Theta}) p(\boldsymbol{Z} \mid \boldsymbol{\Pi}) = \exp(-\mathfrak{L}) \prod_{i=1}^N \prod_{j=0}^{i-1} \left(\lambda_{r_i}^{r_j}(t_i)\right)^{z_{ij}} \prod_{i=1}^N \prod_{j=0}^{i-1} (\pi_{ij})^{z_{ij}} \ , \tag{3}$$

where $\mathfrak{L} = \int_0^T \sum_{r \in \boldsymbol{R}} \lambda_r(t) dt$, $\boldsymbol{\Theta} = \{\boldsymbol{\eta}, \boldsymbol{\xi}, \varphi\}$, and $\boldsymbol{\Pi} = \{\boldsymbol{\pi}_i\}_{i=1}^N$.

The log likelihood of Eq. (3) is represented as

$$\log p(\boldsymbol{D}, \boldsymbol{Z} \mid \boldsymbol{\Theta}, \boldsymbol{\Pi}) = \sum_{i=1}^N \sum_{j=0}^{i-1} z_{ij} \left(\log \lambda_{r_i}^{r_j}(t_i) + \log \pi_{ij}\right) - \mathfrak{L} \ . \tag{4}$$

**Expectation-Maximization.** In order to find the maximum likelihood estimates with latent indices in Eq. (4), we apply the Expectation-Maximization algorithm.

In the E-Step, we calculate the expectation of the joint log likelihood in Eq. (4) given the parameters $\boldsymbol{\Theta}$ and $\boldsymbol{\Pi}$ as

$$\mathbb{E}_{\boldsymbol{Z}|\boldsymbol{D},\boldsymbol{\Theta},\boldsymbol{\Pi}}[\log p(\boldsymbol{D},\boldsymbol{Z}\mid\boldsymbol{\Theta},\boldsymbol{\Pi})] = \sum_{i=1}^{N}\sum_{j=0}^{i-1}\gamma(z_{ij})\left(\log\lambda_{r_i}^{r_j}(t_i) + \log\pi_{ij}\right) - \mathfrak{L} \ , \quad (5)$$

where the responsibility is defined as

$$\gamma(z_{ij}) := \mathbb{E}_{\boldsymbol{Z}|\boldsymbol{D},\boldsymbol{\Theta},\boldsymbol{\Pi}}[z_{ij}] = \frac{\pi_{ij}\,\lambda_{r_i}^{r_j}(t_i;t_j)\exp\left(-\int_{t_j}^{t_i}\lambda_{r_i}^{r_j}(t)\,dt\right)}{\sum_{j=0}^{i-1}\pi_{ij}\,\lambda_{r_i}^{r_j}(t_i;t_j)\exp\left(-\int_{t_j}^{t_i}\lambda_{r_i}^{r_j}(t)\,dt\right)}. \quad (6)$$

In the M-Step, on the other hand, we find the optimal parameters which maximize the expectation of the log likelihood in Eq. (5). We apply gradient descent for $\boldsymbol{\Theta}$ and $\boldsymbol{\Pi}$ as below.

First, we apply gradient descent to find the optimal solution for $\boldsymbol{\Theta} = \{\boldsymbol{\eta}, \boldsymbol{\xi}, \varphi\}$ with the partial derivatives,

$$\frac{\partial\mathbb{E}}{\partial\eta_r} = \sum_{i=1}^{N}\delta(r_i,r)\frac{\gamma(z_{i0})}{\eta_r} - T\rho_r^0 \ , \quad (7)$$

$$\frac{\partial\mathbb{E}}{\partial\xi_r} = \sum_{i=1}^{N}\sum_{j=1}^{i-1}\delta(r_j,r)\frac{\delta(r_j)\gamma(z_{ij})}{\zeta(r_j)} - \sum_{i=1}^{N}\frac{1}{\varphi_{r_i}}\delta(r_i,r)\delta(r_i)\left(1-\kappa\right) \ , \quad (8)$$

$$\frac{\partial\mathbb{E}}{\partial\varphi_r} = -\sum_{i=1}^{N}\sum_{j=1}^{i-1}\delta(r_j,r)\gamma(z_{ij})(t_i-t_j)$$
$$+ \sum_{i=1}^{N}\frac{1}{\varphi_{r_i}^2}\delta(r_i,r)\zeta(r_i)\left(1-\kappa\right) - \sum_{i=1}^{N}\frac{1}{\varphi_{r_i}}\delta(r_i,r)\zeta(r_i)(T-t_i)\kappa \ , \quad (9)$$

where $\kappa := \exp(-\varphi_{r_i}(T-t_i))$.

On the other hand, we introduce Lagrangian multipliers $\lambda_i$ to find the optimal $\boldsymbol{\Pi}$ as

$$\mathbb{E}' = \mathbb{E} - \sum_{i=1}^{N}\lambda_i\left(\sum_{j=0}^{i-1}\pi_{ij} - 1\right) \ . \quad (10)$$

Now, the optimal solution for $\boldsymbol{\Pi}$ should satisfy

$$\frac{\partial\mathbb{E}'}{\partial\pi_{ij}} = \frac{\gamma(z_{ij})}{\pi_{ij}} - \lambda_i = 0 \ , \quad (11)$$

$$\frac{\partial\mathbb{E}'}{\partial\lambda_i} = \sum_{j=0}^{i-1}\pi_j - 1 = 0 \ . \quad (12)$$

Therefore,

$$\pi_{ij} = \frac{\gamma(z_{ij})}{\sum_{j=0}^{i-1}\gamma(z_{ij})} \ . \quad (13)$$

Based on our causal inference method, we construct event networks in which nodes and links represent dengue cases and causal relationships as illustrated in Figure S2. Nodes are color-coded by provided outbreak IDs. A single outbreak ID is considered to be a collection of cascading (or ongoing) local transmission initiated by the same index case. As the figure exhibits, event networks are well clustered according to the outbreak IDs, showing feasible causality inference. This suggests that our inference results can be supportive clues to determine outbreak IDs. In particular, we focus on events in 2003, 2009, and 2013 when the largest dengue outbreaks occurred, as shown in Figure S3.

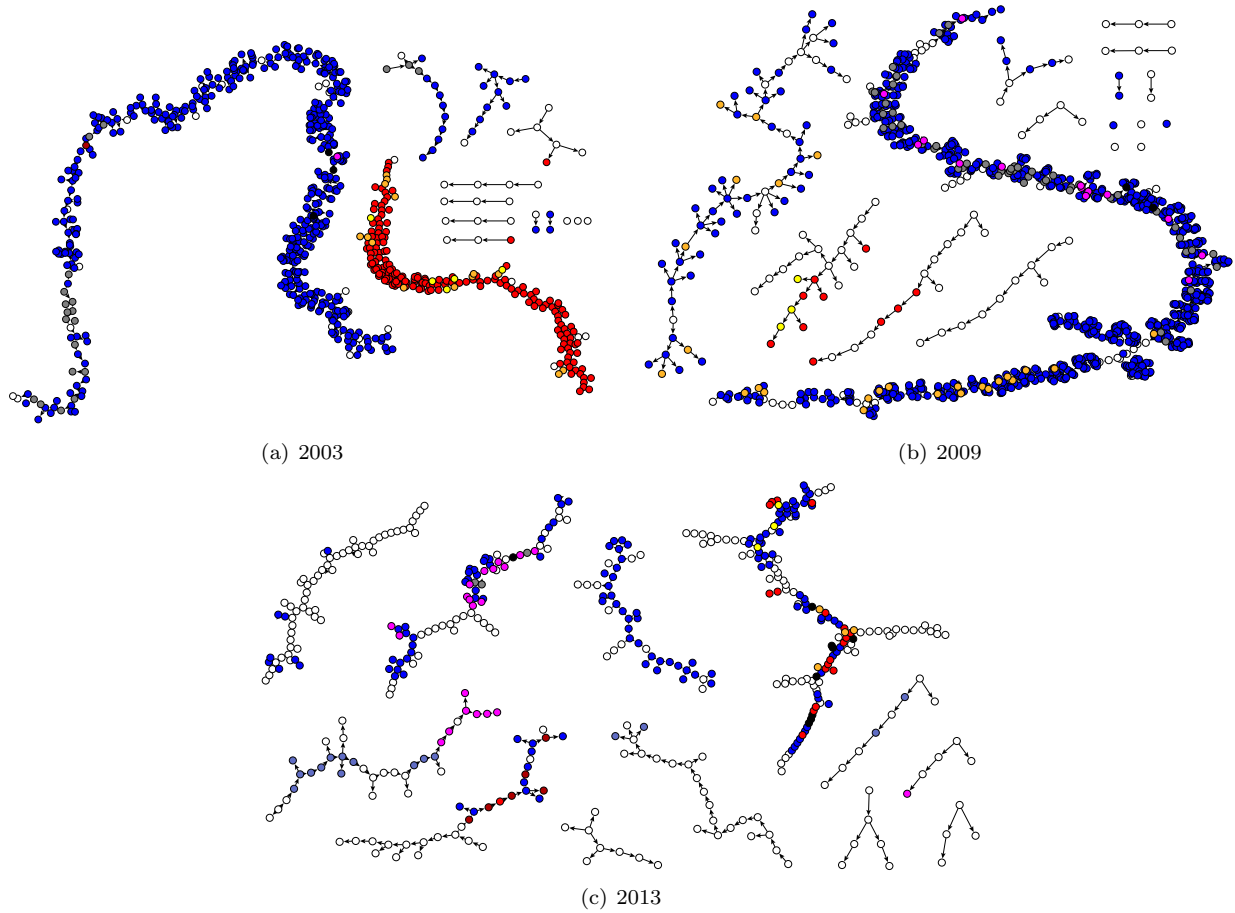

(a) 2003  (b) 2009

(c) 2013

Figure S2: Event networks based on inferred causal relationships between events (dengue cases) for the selected years, (a) 2003, (b) 2009, and (c) 2013, when the largest dengue outbreaks had occurred. Each node and link indicate a dengue case and causality between preceding and triggered events with the highest probability, respectively. Nodes are color-coded by provided outbreak IDs. See Figure 2 in Major Article for the comparison of event networks colored by different regions.

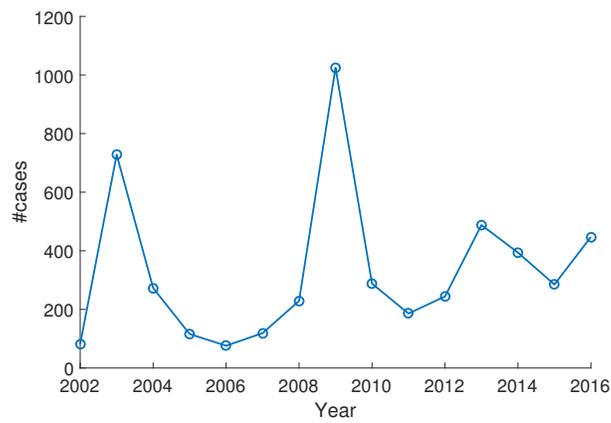

Figure S3: Dengue outbreaks in Queensland between 2002 and 2016.

```
ALGORITHM 1: Simulation of LIPP
   Input  : |R|, η_r, ρ_r^k, ξ_r, φ, T   (1 ≤ r ≤ |R|, 0 ≤ k ≤ |R|)
   Output: D
 1 D ← ∅
 2 t_0 ← 0
 3 i ← 0
 4 while t_i < T do
 5     i ← i + 1
 6     D_i ← ∅
 7     u ← U(0, 1)
 8     for r = 1, ..., |R| do
 9         Δt_{r,i}^0 ← -1/φ log(1 - u)
10         D_i ← D_i ∪ Δt_{r,i}^0
11         for k = 1, ..., |R| do
12             Δt_{r,i}^k ← -1/φ log(1 + φ/λ_r^k(t_{i-1}) log(1 - u))
13             D_i ← D_i ∪ Δt_{r,i}^k
14     t_i ← t_{i-1} + min(D_i)
15     (r*, k*) = arg min_{r,k} D_i
16     D ← D ∪ t_i
17     for r = 1, ..., |R| do
18         for k = 1, ..., |R| do
19             λ_r^k(t_i) ← λ_r^k(t_{i-1}) e^{-φ(t_i - t_{i-1})} + ξ_{r*} ρ_r^{r*}
```

## S3  Simulation

We simulate our proposed model so that we can generate synthetic data as the ground truth and evaluate the model performance. As evaluation metrics, parameter recovery errors are examined with respect to the mean absolute percentage error (MAPE). We also evaluate relative strengths between estimated parameters, which is important to validate the interpretations of underlying diffusion processes with real data. After verifying the proposed model performance with synthetic data, we conduct experiments on real data from epidemics and interpret the dynamics of disease spread as a case study (see Results in Major Article).

### S3.1  Synthetic Data Generation

We describe how to simulate our model LIPP with given parameters. With generated synthetic data, we verify our model by examining how well the model parameters are recovered in terms of parameter value and rank.

**Simulation Algorithm.** Our proposed model is the multi-dimensional Hawkes process, where events are mutually excited across more than one regions. Based on the doubly stochastic property of LIPP, the probability that a new event $i$ arrives within a time interval $\tau$ should consider the whole likelihood of that event generated by the intensity of mutual excitation, $\lambda_r^k(t)$ from an affecting region $k \in \boldsymbol{R}_+$ ($\boldsymbol{R}_+ \equiv \boldsymbol{R} \cup \{0\}$) to a target region $r \in \boldsymbol{R}$ as

$$F_{\Delta t_i}(\tau) = P(\Delta t_i < \tau) = \prod_{r \in \boldsymbol{R}} \prod_{k \in \boldsymbol{R}_+} F_{\Delta t_{r,i}^k}(\tau) \ , \tag{14}$$

where $F_{\Delta t_{r,i}^k}(\tau)$ is the CDF of inter-event time $\Delta t_{r,i}^k = t_{r,i}^k - t_{r,i-1}^k$ produced by the intensity $\lambda_r^k(t)$.

For exogenous infection ($k = 0$), the CDF of its inter-event time $\Delta t_{r,i}^0$ follows

$$F_{\Delta t_{r,i}^0}(\tau) = 1 - \exp\left(\eta_r \ \rho_r^0 \ \tau\right) \ . \tag{15}$$

On the other hand, for the simulation of mutual excitations ($k \neq 0$), we represent the CDF of $\Delta t_{r,i}^k$ as

$$F_{\Delta t_{r,i}^k}(\tau) = 1 - \exp\left(-\frac{1}{\gamma}\lambda_r^k(t_{i-1})\left(1 - e^{-\gamma\tau}\right)\right) \ . \tag{16}$$

We then simulate LIPP by sampling $u$ from a uniform distribution $U(0,1)$ and find the corresponding elapsed time $\Delta t_{r,i}^0$ and $\Delta t_{r,i}^k$ using the inverse transform method [3],

$$\Delta t_{r,i}^0 = -\frac{1}{\gamma}\log(1-u) \ , \tag{17}$$

$$\Delta t_{r,i}^k = -\frac{1}{\gamma}\log\left(1 + \frac{\gamma}{\lambda_r^k(t_{i-1})}\log(1-u)\right) \ , \tag{18}$$

where $\lambda_r^k(t_i) = \lambda_r^k(t_{i-1})e^{-\gamma\Delta t_i} + \xi_k\rho_r^k$ by which we update the intensity for each iteration.

Now, $\Delta t_i$ can be chosen with the minimum inter-arrival time among all $(r \times (k+1))$ samples generated by exogenous and endogenous intensities as

$$\Delta t_i = \min\{\Delta t_{r,i}^0, \Delta t_{r,i}^k\}_{r \in \mathbf{R}, k \in \mathbf{R}+} \ . \tag{19}$$

By sampling both $\Delta t_{r,i}^0$ and $\Delta t_{r,i}^k$ and choosing the minimum $\Delta t_i$, we finally can obtain information about the triggering source of event $i$. This procedure is described in Algorithm 1, and its output is a sequence of timestamps across all regions $\mathbf{R}$.

**Parameter Setting.** We generate 3 groups of synthetic datasets by varying the number of regions such that $|\mathbf{R}| \in \{5, 10, 15\}$, which imitates the number of target regions in our real data. Detailed descriptions on real data will be discussed in the next section. By replicating the estimated parameter values with our real data, we set parameter values for each group such as region $r$'s environmental heterogeneity $\eta_r \in [0.01, 1]$ and latent influence $\xi_r \in [0.5, 2]$. Each group include three different memory kernel exponents such that $\varphi \in \{1, 2, 3\}$, and thus there are 9 subgroups (3 groups × 3 exponents). For each subgroup, we generate 30 cases with random values of $\eta_r$ and $\xi_r$ within the specified ranges above. In total, there are 270 event histories for the synthetic data (9 subgroups × 30 test cases).

### S3.2 Experiments on Synthetic Data

In this section, we verify our proposed model with the simulation data based on parameter recovery errors.

**Parameter Value Recovery.** In order to examine how well our model parameters are recovered from data, we conduct experiments on the three groups of synthetic datasets. Due to the different scale of each parameter, we employ the mean absolute percentage error (MAPE) as a metric for evaluating parameter recovery accuracy as

$$\text{MAPE} = \frac{1}{n}\sum_{i=1}^{n}\left|\frac{A_i - F_i}{A_i}\right| \ , \tag{20}$$

where $A_i$ and $F_i$ represent the actual and forecast parameter values for $n$ synthetic datasets.

Figure S4 depicts the box plot for the estimated parameter errors in MAPE, where parameter values are recovered with about a 40% error on average. That is, our proposed model recovers parameter values with a difference less than one magnitude order compared to the ground truth.

**Parameter Rank Recovery.** However, MAPE cannot show the recovery of relative strengths of parameter values. When it comes to epidemics, the rankings of estimated regional influences are crucial to establish effective strategies. In this regard, we also examine the recovery of relative strengths of true environmental heterogeneity $\boldsymbol{\eta}$ (Figure S5(a)-(c)) and endogenous influence $\boldsymbol{\xi}$ (Figure S5(d)-(f)) for each group ($|\mathbf{R}| \in \{5, 10, 15\}$) by ranking the true ($x$-axis) and its estimated ($y$-axis) parameter values in descending order. Each group has 3 subgroups ($\varphi \in \{1, 2, 3\}$) consisting of 30 test cases, and thus there are 450, 900, and 1,450 rank pairs of true and estimated $\boldsymbol{\eta}$ and $\boldsymbol{\xi}$ for the groups $|\mathbf{R}| \in \{5, 10, 15\}$, respectively. In each plot, the darkness of a grid is proportional to the frequency of the pairs. As the figure shows, shaded grids are overall observed along the line of $y = x$, which shows the order of regional criticality is well recovered. More importantly, the rankings of the most and least influential regions are better recovered, which helps timely and cost effective control on disease spread.

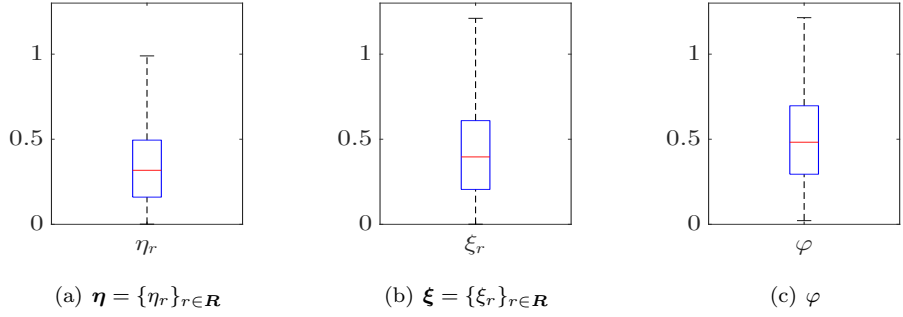

Figure S4: Box plots for the mean absolute percentage errors (MAPE) of estimated parameter values: (a) region $r$'s environmental heterogeneity $\eta_r$, (b) latent influence $\xi_r$, and (c) time decay parameter $\varphi$.

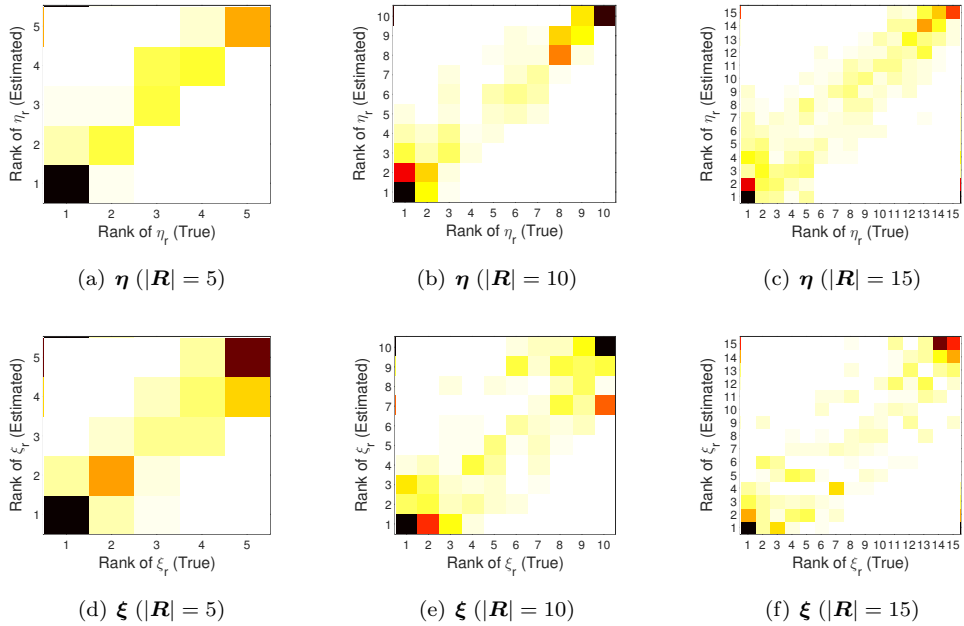

Figure S5: Relative strengths of estimated parameter values compared with the ground truth: (a) to (c) for region $r$'s environmental heterogeneity $\eta_r$ and (d) to (e) for latent influence $\xi_r$. In each plot, $x-$ and $y-$axes represent the pair of true and estimated ranks by varying the number of regions, $|\boldsymbol{R}| \in \{5, 10, 15\}$. The frequency of pairs is color-coded (darker color indicates higher frequency). Shaded grids along the line of $y = x$ imply that the parameter rankings are well recovered.